\documentclass[prd,nofootinbib,showpacs,10pt]{revtex4}
\usepackage{natbib}
\usepackage{amssymb,amsbsy,amsmath,amsfonts}
\usepackage{graphicx}
\usepackage{float}
\usepackage{rotating}

\def\slashp{p \!\!\! \slash}

\begin{document}
\title {Leading-order decuplet contributions to the baryon magnetic moments
in Chiral Perturbation Theory}

\author{L.S. Geng}
\author{J. Martin Camalich}
\author{M.J. Vicente Vacas}
\affiliation{$^1$Departamento de F\'{\i}sica Te\'orica and IFIC, Centro
Mixto Universidad de
Valencia-CSIC,\\ Institutos de Investigaci\'on de Paterna, 46071-Valencia, SPAIN}

\begin{abstract}
We extend an earlier 
study of the baryon magnetic moments in chiral perturbation theory by the explicit inclusion of the spin-3/2 decuplet resonances. We find that the corrections induced by these
heavier degrees of freedom are relatively small in a covariant framework where unphysical spin-1/2 modes are removed. Consequently, implementing the leading SU(3)-breaking corrections given by both the baryon and decuplet contributions, we obtain a description of the baryon-octet magnetic
moments that is better than the Coleman-Glashow relations. Finally, we discuss the uncertainties and compare between  heavy baryon and covariant approaches. 

\end{abstract}

\date{\today}
\maketitle

\section{Introduction}

In the limit that SU(3) is an exact flavor symmetry it is possible to relate the magnetic moments of the baryon-octet and the $\Lambda\Sigma^0$ transition to those of the proton and the neutron. These are the celebrated Coleman-Glashow formulas~\cite{Coleman:1961jn}. The improvement of this description requires the inclusion of a realistic SU(3)-breaking mechanism. The chiral perturbation theory ($\chi$PT)~\cite{Gasser:1984gg,Gasser:1987rb,Scherer:2002tk} is a proper framework to tackle this problem in a systematic fashion.  

In the last decades several calculations of the SU(3)-breaking corrections using $\chi$PT have been performed. It was soon realized that the leading chiral terms overestimate these corrections~\cite{Caldi:1974ta}. Most of the calculations have been done in the context of heavy baryon (HB) $\chi$PT~\cite{Jenkins:1990jv}, with~\cite{Jenkins:1992pi,Durand:1997ya,Puglia:1999th} and without~\cite{Meissner:1997hn} the explicit inclusion of the decuplet resonances. In the HB approach it is necessary to work up to next-to-next-to-leading order (NNLO) to find a good agreement with data, although at the price of the predictiveness of the theory since at this level one has seven unknown low energy constants (LECs) to describe eight measured quantities. Other approaches like cut-off regularized $\chi$PT~\cite{Donoghue:2004vk} and large $N_c$ calculations have been considered~\cite{Luty:1994ub}.

The development of covariant $\chi$PT has been hampered by the problems in the power counting introduced by the baryon mass as a new large scale~\cite{Gasser:1987rb}. Calculations of the baryon magnetic moments in the covariant approach have become available only after the advent of the infrared (IR)~\cite{Becher:1999he} and the extended-on-mass-shell (EOMS)~\cite{Fuchs:2003qc} renormalization schemes. In the IR case~\cite{Kubis:2000aa} at next-to-leading order (NLO) the description is even worse than in the HB approach and it becomes necessary to reach NNLO. On the other hand, it has been found that within the EOMS scheme up to NLO the agreement with the data is not only better than in HB and IR but also than the Coleman-Glashow description~\cite{Geng:2008mf}. In the latter work the differences at NLO among the three different $\chi$PT approaches have been investigated, and the importance of analyticity has been highlighted. 

Nevertheless, in SU(3)-flavor $\chi$PT it is necessary to consider the contributions of the decuplet resonances since the typical scale for their onset $\delta=M_D-M_B\sim0.3$ GeV is well below our expansion parameters $m_K$ and $m_\eta$. The description of higher-spin ($s\geq$3/2) particles in a relativistic field theory is known to be problematic because of the presence of unphysical lower-spin components. For instance, in the Rarita-Schwinger (RS) formulation~\cite{Rarita:1941mf} adopted in this work, the field representation of a massive 3/2-particle is a vector-spinor $\psi_\mu$ with two unphysical spin-1/2 components in addition to the spin-3/2 component. In the presence of interactions the unphysical degrees of freedom are known to lead to pathologies like non-positive definite commutators or acausal propagation for the coupling of the photon~\cite{Johnson:1960vt,Velo:1969bt,Deser:2000dz}. Equivalent problems in phenomenological hadronic interactions have also been largely discussed~\cite{Nath:1971wp,Hagen:1972ea,Singh:1973gq,Pascalutsa:1998pw,Pascalutsa:1999zz}. In the context of $\chi$PT one can use field redefinitions on the conventional chiral Lagrangians in order to cast the interactions in a form that is invariant under the transformation $\psi_\mu\rightarrow\psi_\mu+\partial_\mu\epsilon$~\cite{Pascalutsa:2000kd,Pascalutsa:2006up,Krebs:2008zb}. The resulting gauge symmetry ensures to keep active only the physical degrees of freedom~\cite{Pascalutsa:1998pw}.
Furthermore, there is abundant work concerning the inclusion of spin-3/2 resonances in the framework of baryonic effective field theories~\cite{Tang:1996sq,Hemmert:1997ye,Hacker:2005fh,Wies:2006rv}.

\section{Formalism}
\subsection{Chiral Lagrangians for the baryon-decuplet}
The baryon-decuplet consists of a SU(3)-flavor multiplet of spin-3/2 resonances that we will represent with the Rarita-Schwinger field $T_\mu\equiv T^{ade}_\mu$ with the following associations: 
$T^{111}=\Delta^{++}$, $T^{112}=\Delta^+/\sqrt{3}$,
$T^{122}=\Delta^0/\sqrt{3}$, $T^{222}=\Delta^-$, $T^{113}=\Sigma^{*+}/\sqrt{3}$,
$T^{123}=\Sigma^{*0}/\sqrt{6}$, $T^{223}=\Sigma^{*-}/\sqrt{3}$, 
$T^{133}=\Xi^{*0}/\sqrt{3}$, $T^{233}=\Xi^{*-}/\sqrt{3}$, and $T^{333}=\Omega^-$. The covariantized free Lagrangian is
\begin{equation}
 \mathcal{L}_{D}=\bar{T}^{abc}_\mu(i\gamma^{\mu\nu\alpha}D_\alpha-M_D\gamma^{\mu\nu})T^{abc}_\nu, \label{Eq:RSLag}
\end{equation}
with $M_D$ the decuplet-baryon mass and $D_\nu T_\mu^{abc}=\partial_\nu T_\mu^{abc}+(\Gamma_\nu)_d^a T_\mu^{dbc}
+(\Gamma_\nu)_d^b T_\mu^{adc}+(\Gamma_\nu)_d^c T_\mu^{abd}$. In the last and following Lagrangians we sum over any repeated SU(3)-index denoted by latin characters $a,b,c,\ldots$, and $(X)^a_b$ denotes the element of row $a$ and column $b$ of the matrix representation of $X$. 

The conventional lowest-order chiral Lagrangian for the interaction of the decuplet- and octet-baryons with the pseudoscalar mesons expanded up to one meson field is\footnote{Concerning the building blocks of the chiral Lagrangians, we follow the definitions and conventions of \cite{Scherer:2002tk}}
\begin{eqnarray}
 \mathcal{L}^{(1)}_{\phi BD}=\frac{\mathcal{C}}{F_\phi}\;\varepsilon^{abc}\bar{T}^{ade}_\mu\left(g^{\mu\nu}+z \gamma^\mu\gamma^\nu\right)
 B^e_c\,\partial_\nu\phi^d_b+{\rm h.c.},\label{Eq:CnvLag}
\end{eqnarray}
where $\mathcal{C}$ is the $\phi B D$ coupling, $F_\phi$ the meson-decay constant and $z$ is an off-shell parameter. An analysis of the constraint structure of the interacting theory of Eqs. (\ref{Eq:RSLag}, \ref{Eq:CnvLag}) yields $z=-1$~\cite{Nath:1971wp}. Nevertheless, the resulting interaction leads to well-known problems afflicting the relativistic quantum field theory of 3/2-spinors~\cite{Hagen:1972ea,Singh:1973gq,Pascalutsa:1998pw}.  

The alternative approach of demanding the effective Lagrangians to be spin-3/2-gauge invariant leads, after a field redefinition, to the ``consistent'' $\phi B D$ interaction ~\cite{Pascalutsa:1998pw,Pascalutsa:1999zz}
\begin{equation}
\mathcal{L}\,'^{\,(1)}_{\phi B D}=\frac{i\,\mathcal{C}}{M_D F_\phi}\;\varepsilon^{abc}\left(\partial_\alpha\bar{T}^{ade}_\mu\right)\gamma^{\alpha\mu\nu}
 B^e_c\,\partial_\nu\phi^d_b+{\rm h.c.},\label{Eq:CnsLag}
\end{equation}
which is on-shell equivalent to Eq. (\ref{Eq:CnvLag}). 
Besides, one obtains a second-order $\phi\phi B B$ contact term
\begin{equation}
\mathcal{L}^{(2)}_{\phi\phi B B}=\frac{\mathcal{C}^2}{12M_D^2 F_\phi^2}\left(3\langle \bar{B}\{[\partial_\mu\phi,\partial_\nu\phi],(R^{\mu\nu}B)\}\rangle+\langle \bar{B}[[\partial_\mu\phi,\partial_\nu\phi],(R^{\mu\nu}B)]\rangle-6\langle\bar{B}\partial_\mu\phi\rangle\langle\partial_\nu\phi (R^{\mu\nu}B)\rangle\right)\label{Eq:HOCTLag}
\end{equation}
where $R^{\mu\nu}=i\gamma^{\mu\nu\alpha}\partial_\alpha+M_D\gamma^{\mu\nu}$ and $\langle\ldots\rangle$ denotes the trace in flavor space. The latter Lagrangian is interpreted as carrying the spin-1/2 content of the Lagrangian (\ref{Eq:CnvLag}). This term is eliminated by absorbing it into suitable higher-order LECs.

\subsection{Power Counting}

We use the standard power counting where one assigns the chiral order $n_{\chi PT}=4L-2N_M-N_B+\sum_k k V_k$ to a diagram with $L$ loops, $N_M$ ($N_B$) internal meson (octet- and decuplet-baryon) propagators and $V_k$ vertices from $k$th order Lagrangians. In the covariant theory with the modified minimal subtraction method ($\overline{MS}$), this rule is violated by lower-order analytical pieces. In order to recover the power counting we apply the EOMS renormalization prescription~\cite{Fuchs:2003qc}. For the diagrams with internal decuplet-baryon lines we absorb into the LECs the terms breaking the power counting that are obtained expanding the loop-functions around the chiral limit. 
 
Besides, the propagator corresponding to the RS action in $d$ dimensions
\begin{equation}
S^{\mu\nu}(p)=-\frac{\slashp+M_D}{p^2-M_D^2+i\epsilon}\left[
g^{\mu\nu}-\frac{1}{d-1}\gamma^\mu\gamma^\nu-\frac{1}{(d-1)M_D}\left(\gamma^\mu\, p^\nu-\gamma^\nu\, p^\mu\right)
-\frac{d-2}{(d-1)M_D^2} p^\mu p^\nu \right],
\end{equation}
has a problematic high-energy behavior~\cite{Piccinini:1984dd}. In the context of an effective field theory, this is responsible for the appearance of $d$ - 4 singularities of a chiral order higher than the one naively expected using the power counting. These infinities would be absorbed by the proper counter-terms to be included at next orders. However, we do not include these terms explicitly but perform a $\overline{MS}$-subtraction on them and study the residual regularization-scale dependence.

\subsection{Parameter values}

The $\phi B D$ coupling $\mathcal{C}=\frac{h_A}{2\sqrt{2}}\approx1.0$ is obtained by fitting $\Delta\rightarrow N\pi$ decay
width~\cite{Pascalutsa:2006up}\footnote{Note that the value for $\mathcal{C}$ of the present work is different from the one often used in HB calculations~\cite{Jenkins:1992pi,Durand:1997ya,Puglia:1999th}. In these papers, it is applied a convention for the ``vielbein'' that is related to ours by a factor of 2. Consequently, the values in the HB studies, $\mathcal{C}\sim$1.5, equal $\mathcal{C}\sim$0.75 in our convention. Notice also the corresponding differences in the coefficients in Table \ref{Table:Coefficients}.}. In the SU(3) context, the value of $\mathcal{C}$ can vary depending on
the decay channel, with the one obtained from the $\Delta\rightarrow N\pi$ decay
being the largest ~\cite{Butler:1992pn}. This fact is effectively taken into account in
our present study by using an average $F_\phi\equiv1.17f_\pi$ with $f_\pi=92.4$ MeV. 
Therefore, we use $\mathcal{C}=1$ in our present study unless otherwise stated. For the masses of the pseudoscalar mesons we take
$m_\pi\equiv m_{\pi^\pm}=0.13957$ GeV, $m_K\equiv m_{K^\pm}=0.49368$ GeV, 
$m_\eta=0.5475$ GeV while for the baryon masses we use the average among the members of the respective SU(3)-multiplets, $M_B=1.151$ GeV and $M_D=1.382$ GeV. A moderate variation of $M_B$
and $M_D$ is investigated below. 

\begin{figure}[t]
\includegraphics[width=\columnwidth]{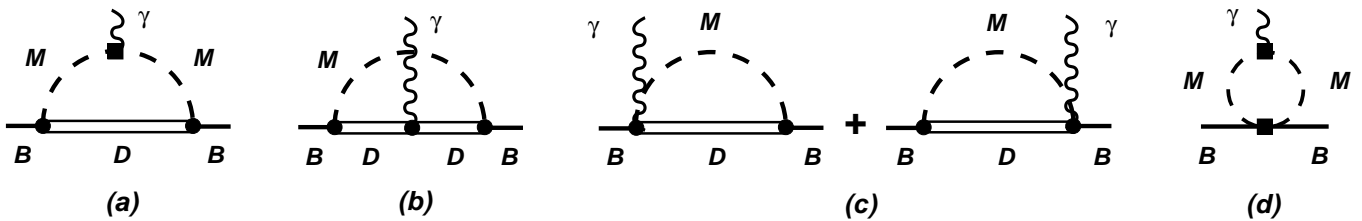}
\caption{Feynman diagrams with internal decuplet-resonances studied in this work. The solid lines correspond to octet-baryons, the double lines to decuplet-resonances, the dashed lines to mesons and the wiggly line denotes the external photon field. Black dots and boxes indicate $\mathcal{O}(p)$ and $\mathcal{O}(p^2)$ couplings respectively. Diagrams \textbf{\textit{(a)}}-\textbf{\textit{(c)}} contribute at $\mathcal{O}(p^3)$, while diagram \textbf {\textit{(d)}} is a  $\mathcal{O}(p^4)$-contribution that represents the difference between the calculation with consistent couplings and with conventional ones where $z=-1$ (see section III).\label{Fig:diagram}}
\end{figure}

\section{Results}

The Feynman diagrams with internal decuplet resonances that contribute to the baryon-octet anomalous magnetic moments at $\mathcal{O}(p^3)$ are those of Fig. \ref{Fig:diagram} \textbf{\textit{(a)}}-\textbf{\textit{(c)}} . Their contribution is to be added to the ones coming from the SU(3)-symmetric LECs $b_6^D$ and $b_6^F$, and from the diagrams with internal octet-baryons~\cite{Geng:2008mf}. The calculation of the decuplet diagrams with the consistent couplings of Eq. (\ref{Eq:CnsLag}) results in
\begin{equation}
\tilde{\kappa}_B^{(3)}= \frac{\mathcal{C}^2M_D^2}{8\pi^2 F_\phi^2}\big(\sum_{M=\pi,K}\tilde{\xi}^{(a)}_{BM}\big(H^{(a)}(m_M,\mu)-H^{(c,I)}(m_M,\mu)\big)+\sum_{M=\pi,K,\eta}\tilde{\xi}^{(b)}_{BM}\big(H^{(b)}(m_M,\mu)-H^{(c,II)}(m_M,\mu)\big)\big),\label{Eq:DResult}
\end{equation}
with the coefficients $\tilde{\xi}^{(a,b)}$ listed in Table \ref{Table:Coefficients} (Appendix). The loop functions $H^{(X)}$ can be found in the Appendix in terms of Feynman-parameter integrals. The additional character that appears in the loop-functions of the diagram \textbf{\textit{(c)}} indicates whether the seagull-diagram comes from the minimal substitution performed in the derivative of the meson fields ($I$) or of decuplet fields ($II$) in the consistent approach of Eq. (\ref{Eq:CnsLag}). 

In order to recover the power counting we apply the EOMS-scheme. After removing the $\mathcal{O}(p^2)$ ultraviolet divergences by the $\overline{MS}$ procedure, this is equivalent to redefine the two LECs obtained in \cite{Geng:2008mf} as
\begin{equation}
\hat{b}_6^D=\tilde{b}_6^D+\frac{\mathcal{C}^2M_D^2}{8 \pi^2F_\phi^2}f^D(\mu),\hspace{1cm}\hat{b}_6^F=\tilde{b}_6^F+\frac{\mathcal{C}^2M_D^2}{8 \pi^2F_\phi^2}f^F(\mu),
\end{equation}
where the functions $f^{D,F}(\mu)$ can be found in the Appendix. These functions, as well as the loop-functions, depend on the regularization-scale $\mu$. We use a value $\mu=1$ GeV and analyze below our results with respect to a moderate variation around this value. From the EOMS-renormalized loop-functions $\hat{H}^{(X)}$ one can then obtain the non-analytical pieces of the HB results assuming $M_D=M_B+\delta$ and applying that $M_B\sim\Lambda_{\chi SB}$ in what nowadays is known as the $\epsilon$-expansion~\cite{Hemmert:1997ye}. One finds that at NLO the only non-zero contribution comes from diagram \textbf{\textit{(a)}}, giving
\begin{eqnarray}
M_D^2\,\hat{H}^{(a)}(m)\simeq-\delta\,M_B\log\left(\frac{m^2}{4\delta^2}\right)+\left\{\begin{array}{c}
                             2\,M_B\sqrt{m^2-\delta^2}\left(\frac{\pi}{2}-\arctan\left(\frac{\delta}{\sqrt{m^2-\delta^2}}\right)\right)\hspace{1.3cm}m\geq\delta \nonumber\\
M_B\sqrt{\delta^2-m^2}\log\left(\frac{\delta-\sqrt{\delta^2-m^2}}{\delta+\sqrt{\delta^2-m^2}}\right)\hspace{2.8cm}m<\delta
                             \end{array}\right.,
\end{eqnarray}
in agreement with the result of~\cite{Jenkins:1992pi}.
Moreover, one obtains the decoupling of the decuplet resonances for the limit where $M_D\rightarrow\infty$. Indeed, one finds that the EOMS-renormalized loop-functions verify $\lim_{M_D\to\infty}\,\hat{H}^{(X)}=0$. 

We have also done the calculation using the conventional couplings of Eq. (\ref{Eq:CnvLag}) with the choice $z=-1$. We found that the results of the seagull \textit{\textbf{(c)}} of type $I$ are the same in both approaches, whereas the diagram \textit{\textbf{(b)}} with conventional couplings equals the same diagram \textit{\textbf{(b)}}  minus the seagull \textit{\textbf{(c)}} of type $II$ in the case of consistent couplings. The only difference between both calculations comes from diagram \textit{\textbf{(a)}}. If we subtract the corresponding loop-function obtained using consistent couplings $H^{(a)}$ from the one obtained with the conventional couplings $H'^{(a)}$ we find  
\begin{equation}
\delta H^{(a)}(m,\mu)=H'^{(a)}(m,\mu)-H^{(a)}(m,\mu)=\frac{3 M_B(M_B+M_D)}{2M_D^4}\;m^2\left(1-\log\left(\frac{m^2}{\mu^2}\right)\right).\label{Eq:HOTadpole}
\end{equation}
One can check that this is the contribution given by the tadpole Fig \ref{Fig:diagram}-\textit{\textbf{(d)}} where the $\phi\phi BB$ vertex is the one obtained from Eq. (\ref{Eq:HOCTLag}). Therefore, Eq. (\ref{Eq:HOTadpole}) is the higher-order contribution to the anomalous magnetic moments of the octet-baryons that is removed when using consistent couplings and interpreted as coming from unphysical degrees of freedom. Indeed, the difference between both approaches comes from diagram \textit{\textbf{(a)}} for which the consistent couplings eliminate completely the spurious spin-1/2 components. On the other hand, both schemes include the non-consistent minimal $\gamma DD$ coupling. In this regard, we observe that the loop-contribution with this coupling gives the same result in both frameworks. Finally, notice also that $\lim_{M_D\to\infty}\,\delta H^{(a)}$=0.\footnote{In the calculations done in this work the electromagnetic gauge invariance have been checked. We have computed the  loop-contributions (Fig. \ref{Fig:diagram}) to the electric charge of any octet-baryon $\delta Q_B$ and have verified that they are canceled by the wave-function renormalization  $\Sigma'_B$ of the minimal photon coupling: $\delta Q_B+Q_B\,\Sigma'_B=0$.}

\begin{table*}
\centering
\caption{Baryon octet magnetic moments in chiral perturbation theory up to $\mathcal{O}(p^3)$. We compare the SU(3)-symmetric description with the different $\mathcal{O}(p^3)$ $\chi$PT calculations discussed in the text. Namely, we display the Heavy Baryon and the covariant-EOMS results both with (O+D) and without (O) the inclusion of dynamical decuplets. In the covariant case we show the numerical results obtained using the consistent couplings (\ref{Eq:CnsLag}) and the conventional couplings (\ref{Eq:CnvLag}) with $z=-1$. We also include the experimental values for reference~\cite{Amsler:2008zzb}.  \label{Table:Results}}
\begin{tabular}{c|c|cc|ccc|c|}
\cline{2-8}
& &\multicolumn{2}{|c|}{Heavy Baryon $\mathcal{O}(p^3)$}&\multicolumn{3}{|c|}{Covariant EOMS $\mathcal{O}(p^3)$}& \\
\cline{3-7}
&  \raisebox{1ex}[0pt]{Tree level $\mathcal{O}(p^2)$}& O & O+D& O & O+D (conv.) &O+D (consist.) & \raisebox{1ex}[0pt]{Expt.}\\ 
\hline\hline
\multicolumn{1}{|c|}{\textit{p}} & 2.56 &3.01& 3.47 & 2.60 & 3.18 & 2.61& 2.793(0)\\
\multicolumn{1}{|c|}{\textit{n}} & -1.60 &-2.62&-2.84& -2.16 & -2.51&-2.23 &-1.913(0)\\
\multicolumn{1}{|c|}{$\Lambda$}& -0.80 &-0.42&-0.17&-0.64&-0.29 &-0.60&-0.613(4) \\
\multicolumn{1}{|c|}{$\Sigma^-$}& -0.97 &-1.35&-1.42& -1.12 &-1.26&-1.17&-1.160(25)\\
\multicolumn{1}{|c|}{$\Sigma^+$}& 2.56 &2.18&1.77& 2.41& 1.84 &2.37&2.458(10) \\
\multicolumn{1}{|c|}{$\Sigma^0$}& 0.80 &0.42&0.17& 0.64 & 0.29& 0.60 & ... \\
\multicolumn{1}{|c|}{$\Xi^-$}& -1.60 &-0.70&-0.41& -0.93 & -0.78& -0.92 & -0.651(3) \\
\multicolumn{1}{|c|}{$\Xi^0$}&-0.97 &-0.52&-0.56& -1.23 & -1.05& -1.22 & -1.250(14) \\
\multicolumn{1}{|c|}{$\Lambda\Sigma^0$} & 1.38 &1.68&1.86& 1.58 & 1.88 & 1.65& $\pm$1.61(8)\\
\hline
\multicolumn{1}{|c|}{$b_6^D$}& 2.40&4.71&5.88 & 3.92 & 5.76 & 4.30 & \\
\multicolumn{1}{|c|}{$b_6^F$}& 0.77 &2.48&2.49& 1.28 & 1.03 & 1.03 & ...\\
\multicolumn{1}{|c|}{$\bar{\chi}^2$}&0.46&1.01&2.58& 0.18 & 1.06 & 0.22& \\
\hline
\end{tabular}
\end{table*}

In Table \ref{Table:Results} we show the numerical results for the baryon magnetic moments obtained by minimizing $\bar{\chi}^2=\sum(\mu_{th}-\mu_{expt})^2$ as a function of the LECs $b_6^D$ and $b_6^F$ renormalized as described before. We have not included the $\Lambda\Sigma^0$ transition moment in the fit and, therefore, it is a prediction. We compare the SU(3)-symmetric description and different $\chi$PT approaches providing the leading breaking corrections. Namely, we display the HB and the covariant-EOMS results both with (O+D) and without (O) the inclusion of dynamical decuplets. In the covariant case we show the numerical results obtained using the consistent couplings (\ref{Eq:CnsLag}) and the conventional couplings (\ref{Eq:CnvLag}) with $z=-1$. We also include the experimental values for reference~\cite{Amsler:2008zzb}. 

For the HB approach, one sees how the corrections of the dynamical baryon -octet and -decuplet go in the same direction and are of equivalent size. Consequently, the description obtained with only the baryon-octet, that already overestimated the SU(3)-breaking corrections, gets much worsened. In the covariant case we obtain two quite different results depending on whether we use the consistent or the conventional ($z=-1$) couplings. For the latter, we find that in general the corrections given by the decuplet resonances are quite large and tend to spoil the NLO improvement over the Coleman-Glashow description. 

In the covariant formulation with consistent couplings, the decuplet contributions are small and added to the octet contributions provide an overall description of the same quality as that obtained with only octet-baryons within EOMS. In this case, we can study the convergence properties of the chiral series factorizing the tree-level at $\mathcal{O}(p^2)$ from the whole result up to $\mathcal{O}(p^3)$. We also separate the loop fraction into the octet (second number) and the decuplet (third number) parts in the parenthesis
\begin{eqnarray*}
&&\mu_p=3.46(1-0.28+0.035)\hspace{0.15cm},\hspace{0.15cm}\mu_n=-2.86(1-0.16-0.06)\hspace{0.15cm},\hspace{0.15cm}\mu_\Lambda=-1.43(1-0.46-0.12),\\
&&\mu_{\Sigma^-}=-0.60(1+0.25+0.70)\hspace{0.15cm},\hspace{0.15cm}\mu_{\Sigma^+}=3.46(1-0.34+0.025)\hspace{0.15cm},\hspace{0.15cm}\mu_{\Sigma^0}=1.41(1-0.47-0.11),\\
&&\mu_{\Xi^-}=-0.60(1-0.07+0.61)\hspace{0.15cm},\hspace{0.15cm}\mu_{\Xi^0}=-2.86(1-0.48-0.09)\hspace{0.15cm},\hspace{0.15cm}\mu_{\Lambda\Sigma^0}=2.48(1-0.28-0.06).\label{Eq:convergence}
\end{eqnarray*}
Except for the $\Sigma^-$, the relative contributions of the octet and the decuplet and the overall $\mathcal{O}(p^3)$ corrections, are consistent with a maximal correction of about $m_\eta/\Lambda_{\chi SB}$.

 Among the set of sum-rules obtained by Caldi and Pagels~\cite{Caldi:1974ta} two of them survive up to the leading breaking corrections provided by any of the covariant $\chi$PT approaches considered. Namely, we found that our results verify
\begin{equation}
 \mu_{\Sigma^+}+\mu_{\Sigma^-}=-2\mu_\Lambda,\hspace{1cm}\mu_{\Lambda\Sigma^0}=\frac{1}{\sqrt{3}}\left(\mu_\Lambda-\mu_{\Xi^0}-\mu_n\right). \label{Eq:C-PSum-Rules}
\end{equation}
The first relation in combination with the assumed isospin symmetry is the cause of $\mu_\Lambda=-\mu_{\Sigma^0}$ in the results of Table \ref{Table:Results}. Experimentally, the two relations in Eq. (\ref{Eq:C-PSum-Rules}) are satisfied rather accurately, 1.298(27)=1.226(8) for the first relation and 1.61(8)=1.472(8) for the second. A combination of them produces the Okubo sum-rule~\cite{Okubo:1963zza}. The third sum-rule derived in~\cite{Caldi:1974ta}, although fulfilled in the HB expansions of our results (see Ref.~\cite{Jenkins:1992pi}), is broken when the relativistic corrections to the loops are included. 

The comparison between the results of the three approaches to implement the decuplet resonances presented in this work deserves some comments. The covariant framework provides a much better description of the resonances effects than the HB one. For the results obtained in the latter, one is unavoidably led to wonder about the contributions of higher-mass resonances. In the covariant formulation, relativistic corrections in form of higher-order terms in the expansion on $m/\Lambda_{\chi SB}$ and $\delta/ \Lambda_{\chi SB}$, are resummed in a way that preserves analyticity. However, in the covariant approach one faces the problem of the spurious degrees of freedom and their unphysical contributions. We identified above the term that produces the difference between the two covariant calculations, see Eqs. (\ref{Eq:HOCTLag}) and (\ref{Eq:HOTadpole}). Moreover, we interpreted this difference in the context of the spin-3/2 gauge symmetry as a contribution of the spin-1/2 modes that are decoupled after applying a suitable field redefinition. In conclusion, the  differences exhibited in Table \ref{Table:Results} highlight the importance of settling a proper framework to implement the spin-3/2 resonances into an effective field theory.

\begin{figure}[t]
\includegraphics[width=\columnwidth]{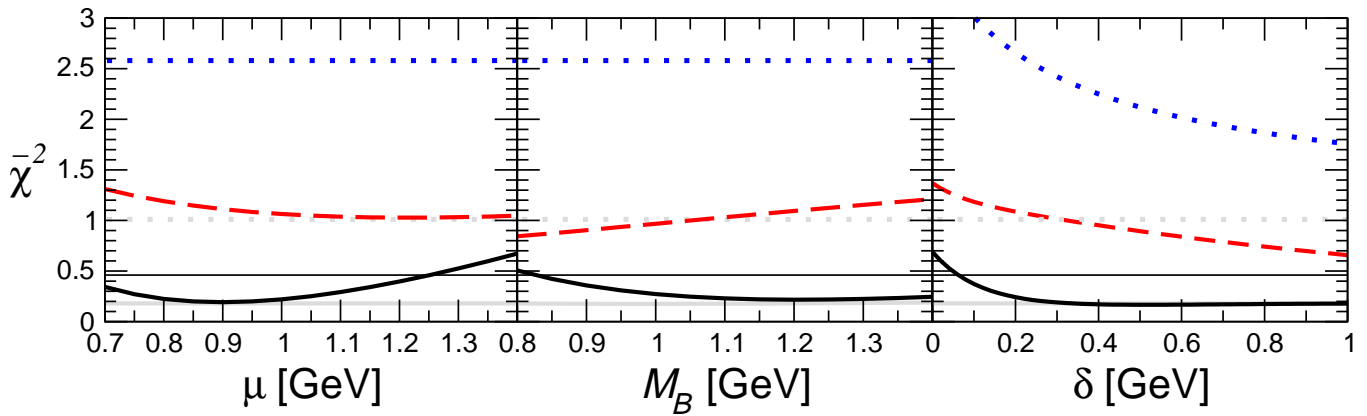}
\caption{Uncertainties of the numerical results of Table \ref{Table:Results} due to the values of the regularization scale $\mu$ (\textit{left} panel), the average baryon mass $M_B$ (\textit{center} panel) and the decuplet-octet mass splitting $\delta$ (\textit{right} panel). The lines represent the $\bar{\chi}^2$ for the results obtained in HB (dotted line) and in covariant approach using conventional (dashed line) and consistent (thick solid line) couplings. The grey lines represent the $\bar{\chi}^2$ for the case without explicit decuplet resonances in HB (dotted) and in covariant formulation (solid). For reference we also include the SU(3)-symmetric description (thin solid line).  \label{Fig:Graphs}}
\end{figure}

In Fig. \ref{Fig:Graphs} we collect the uncertainties of the numerical results due to the values of the parameters used in this work. The graphs represent the dependence of $\bar{\chi}^2$ on the regularization scale $\mu$ (\textit{left}), the baryon mass $M_B$ (\textit{center}) and the decuplet-octet mass splitting $\delta$ (\textit{right}) in HB (dotted line) and in the covariant formulation using conventional (dashed line) and consistent (thick solid line) couplings. We also show the $\bar{\chi}^2$ of the results without decuplet resonances in grey (solid and dotted for covariant and HB respectively) and of the SU(3)-symmetric description (horizontal thin solid line). The three graphs show that the results given in Table \ref{Table:Results} are representative of those obtained for any other values of the parameters $\mu$, $M_B$ and $\delta$ chosen within reasonable intervals. In particular, the first graph shows that the covariant calculation with consistent couplings improves the SU(3)-symmetric description for 0.7 GeV $\leq\mu\leq$ 1.3 GeV, being the best description for $\mu\sim M_N\simeq$0.94 GeV. From the second graph we conclude that the results are solid with respect to a variation of the average baryon mass and keeping $\delta=0.231$ GeV. We have observed numerically that the size of the decuplet contributions in the covariant case with consistent couplings decreases as  $\delta$ increases and practically reaches the decoupling for $\delta\sim0.3$ GeV. A manifestation of this can be seen in the \textit{right} panel. 

Finally we want to make also some comments on the problem of non-consistency of the minimal coupling of the photon to the RS-fields. There is not an accepted solution to this problem although it has been argued that the inclusion of a set of higher-order non-minimal terms could improve the situation by making the RS field to fulfill low-energy unitarity and the constraint analysis in an approximated way~\cite{Deser:2000dz}. To have some insight into the uncertainties that can be brought by the lack of consistency of the minimal coupling, we have added to our covariant results the contribution of diagram \textbf{\textit{(b)}} of Figure \ref{Fig:diagram} with the aforementioned non-minimal $\gamma DD$ coupling~\cite{Deser:2000dz}. We find that the results do not change much, producing even a little improvement over those presented in Table \ref{Table:Results} . For instance, for the consistent couplings, the $\bar{\chi}^2$ is reduced down to $\sim$0.20 and the values for the different magnetic moments change by less than 5\%. The major effect is to reduce the scale dependence, enlarging the range of improvement over the Coleman-Glashow description to the interval of 0.5 GeV $\leq\mu\leq$ 1.8 GeV.

Further progress in the understanding of the baryon octet magnetic moments in $\chi$PT can be straightforwardly addressed at NNLO, although at the cost of the introduction of several new LECs that will harm the predictive power of the theory. Another promising line of work is the description of the electromagnetic structure of the decuplet resonances and of the decuplet-octet transitions using the covariant $\chi$PT approach described here and in Ref.~\cite{Geng:2008mf}.

In summary, we have studied the leading SU(3)-breaking contributions to the baryon-octet magnetic moments produced by the explicit inclusion of the decuplet resonances in chiral perturbation theory. Special attention has been given to the comparison of results obtained using different descriptions of the spin-3/2 resonances. We have shown that the dependence of our results on the renormalization scale $\mu$, the average baryon mass $M_B$, and the decuplet-octet mass splitting is quite mild. Other aspects like the convergence of the chiral series, the decoupling of the decuplet resonances, and the validity of several SU(3) sum rules have also been examined. Particularly, we find that the improvement over the Coleman-Glashow description obtained with only octet-baryons~\cite{Geng:2008mf} remains essentially unchanged after the proper inclusion of the decuplet resonances.  
\section{Acknowledgments}

The authors thank J. Gegelia and V. Pascalutsa for useful discussions. This work was partially supported by the  MEC grant  FIS2006-03438 and the European Community-Research Infrastructure
Integrating Activity Study of Strongly Interacting Matter (Hadron-Physics2, Grant Agreement 227431) under the Seventh Framework Programme of EU. L.S.G. acknowledges support from the MICINN in the Program 
``Juan de la Cierva''. J.M.C. acknowledges the same institution for a FPU grant. 
\section{Appendix}
In the calculation of the loop diagrams, we have used the following
$d$-dimensional integrals in Minkowski space:
\begin{equation}
\int d^d k\frac{k^{\alpha_1}\ldots k^{\alpha_{2n}}}{(\mathcal{M}^2-k^2)^\lambda}
=i\pi^{d/2}\frac{\Gamma(\lambda-n+\varepsilon-2)}{2^n\Gamma(\lambda)}\frac{(-1)^n g^{\alpha_1
\ldots\alpha_{2n}}_s}{(\mathcal{M}^2)^{\lambda-n+\varepsilon-2}}
\end{equation}
with $g^{\alpha_1 \ldots\alpha_{2n}}_s=g^{\alpha_1\alpha_2}\ldots g^{\alpha_{2n-1}\alpha_{2n}}+\ldots$
a combination symmetrical with respect to the permutation of any pair of indices (with $(2n-1)!!$
terms in the sum). We will present the divergent part of the loops as the contact piece $\lambda_\varepsilon=2/\varepsilon+\log{4\pi}-\gamma_E$, where $\varepsilon=4-d$ and $\gamma_E\simeq0.5772$. 

We display below the loop-functions  $H^{(X)}$ of the diagrams of Fig. \ref{Fig:diagram} with $M_B=r\,M_D$, $\mu=\bar{\mu}M_D$ and defining $\mathcal{M}=(1-x)m^2+xM_D^2-x(1-x)M_B^2$, $\mathcal{M}=\bar{\mathcal{M}} M_D^2$
\begin{eqnarray}
&&H^{(a)}=\frac{r^2}{6}\int^1_0\,dx\,(1-x) \,\Big((r (r (1-x)+1) (11 x-2)-14 \bar{\mathcal{M}})+\nonumber\\
&&3\left(\left(1+x-2 x^2\right) r^2+(2 x+1)
   r+4 \bar{\mathcal{M}}\right)\left(\lambda_\varepsilon+\log
   \left(\frac{\bar{\mathcal{M}}}{\bar{\mu} ^2}\right)\right)\Big),\label{Eq:LF-a}
\end{eqnarray}
\begin{eqnarray}
&&H^{(b)}=\frac{r^2}{18} \int^1_0\,dx\,x\,\Big(r (-36 x-r (x-1) (r (8 r (x-1)-3 x+30) (x-1)+9 x-42)+38)+42-\nonumber\\
&&\left(r (50 r (x-1)+15 x+4)-48\right)\bar{\mathcal{M}}+\Big(3 (r (x-1)-1)^2 (r (2 r (x-1)+3 x-2)+6)+\nonumber\\
&&6 (r (4 r (x-1)+6 x-13)-15)\bar{\mathcal{M}}\Big)\left(\lambda_\varepsilon+\log
   \left(\frac{\bar{\mathcal{M}}}{\bar{\mu} ^2}\right)\right)\Big),\label{Eq:LF-b}
\end{eqnarray}
\begin{equation}
H^{(c,I)}=\frac{1}{3}\int^1_0\,dx\,r (1+r (1-x)) \bar{\mathcal{M}} \left(6\left(\lambda_\varepsilon+\log
   \left(\frac{\bar{\mathcal{M}}}{\bar{\mu}^2}\right)\right)-1\right),\label{Eq:LF-cI}
\end{equation}
\begin{equation}
 H^{(c,II)}=-\frac{1}{3}\int^1_0\,dx\,r (1+r(1-x)) \bar{\mathcal{M}} \left(3 (2-r)\left(\lambda_\varepsilon+\log
   \left(\frac{\bar{\mathcal{M}}}{\bar{\mu} ^2}\right)\right)-1-r\right).\label{Eq:LF-cII}
\end{equation}

The functions $f^D$ and $f^F$ used in the regularization of the loop-functions Eqs. (\ref{Eq:LF-a}-\ref{Eq:LF-cII}) are
\begin{eqnarray}
&&f^D(\mu)=\frac{1}{36 r^3}\Big(4 (r (r (r (r+4)+2)-31)-36) \log (\mu ) r^5+\nonumber\\
&&(r (r (r (2 r
   (r (2 r (2 r+5)+9)-49)-
161)+22)+22)-8) r^2-\nonumber\\
&&2 (r+1)^3 \left(r
   \left(r \left(r \left(r
   \left(r^2+r-4\right)-23\right)+44\right)-23\right)+4\right)
   \log \left(1-r^2\right)\Big),\\
&&f^F(\mu)=\frac{5}{108 r^4}\Big(4 (r (r+2) (r (r (r+2)+7)-18)-36) \log (\mu ) r^5+\nonumber\\
&&(r (r (r
   (r (r (r (4 r (2 r+5)+39)-8)-125)-44)+49)-8)-18) r^2-\nonumber\\
&&2
   (r-1)^2 (r+1)^4 (r (r (r (r+2)+8)-14)+9) \log
   \left(1-r^2\right)\Big).
\end{eqnarray}

\begin{table*}
\caption{Coefficients of the loop contributions [Eq. (\ref{Eq:DResult})] to the magnetic moments of the octet baryons.\label{Table:Coefficients}}

\centering
\begin{tabular}{cccccccccc}
\hline\hline 
 & $p$ & $n$ & $\Lambda$ & $\Sigma^-$ & $\Sigma^+$ & $\Sigma^0$ & $\Xi^-$ & $\Xi^0$ & $\Lambda\Sigma^0$  \\ 
\hline
$\tilde{\xi}^{(a)}_{B\pi}$ &  $-\frac{8}{9}$ & $\frac{8}{9}$ & 0 & $-\frac{2}{9}$ & $\frac{2}{9}$ & 0 & $-\frac{4}{9}$ & $\frac{4}{9}$ & $-\frac{4}{3\sqrt{3}}$ \\

$\tilde{\xi}^{(a)}_{B K}$ & $\frac{2}{9}$ & $\frac{4}{9}$ & $\frac{2}{3}$ & $-\frac{4}{9}$ & $-\frac{8}{9}$ & $-\frac{2}{3}$ & $-\frac{2}{9}$ & $\frac{8}{9}$ & $-\frac{2}{3\sqrt{3}}$\\

$\tilde{\xi}^{(b)}_{B\pi}$ & $\frac{32}{9}$ & $-\frac{8}{9}$ &  0  & $-\frac{2}{9}$ & $\frac{2}{9}$ &  0 & $-\frac{2}{9}$ & $-\frac{4}{9}$ & $\frac{4}{3\sqrt{3}}$ \\

$\tilde{\xi}^{(b)}_{B K}$ & $\frac{4}{9}$ & $ -\frac{4}{9}$ & $-\frac{2}{3}$ & $-\frac{16}{9}$ & $ \frac{28}{9}$ & $ \frac{2}{3}$ &  $-\frac{16}{9}$ &  $ -\frac{8}{9}$ &  $\frac{2}{3\sqrt{3}}$\\

$\tilde{\xi}^{(b)}_{B \eta}$ & 0 &  0 &  0 & $-\frac{2}{3}$ & $\frac{2}{3}$ &  0 & $-\frac{2}{3}$ &  0 &   0 \\
\hline \hline 
\end{tabular}
\end{table*}
 
\end{document}